\documentclass[pra,twocolumn,unsortedaddress,superscriptaddress]{revtex4}
\usepackage{graphicx}
\usepackage{latexsym,epsfig,graphicx}
\usepackage{dcolumn}
\usepackage{bm}
\usepackage[colorlinks,urlcolor=blue,citecolor=blue]{hyperref}
\usepackage{comment}
\usepackage{lipsum}

\begin{document}

\title{Time-reversal-invariant spin-orbit-coupled bilayer Bose-Einstein
Condensates}
\author{Matthew Maisberger}
\thanks{Authors with equal contribution.}
\affiliation{Department of Physics, The University of Texas at Dallas, Richardson, Texas
75080-3021, USA}
\author{Lin-Cheng Wang}
\thanks{Authors with equal contribution.}
\affiliation{Department of Physics, The University of Texas at Dallas, Richardson, Texas
75080-3021, USA}
\affiliation{School of Physics, Dalian University of Technology, Dalian 116024, PR China}
\author{Kuei Sun}
\affiliation{Department of Physics, The University of Texas at Dallas, Richardson, Texas
75080-3021, USA}
\author{Yong Xu}
\affiliation{Department of Physics, The University of Texas at Dallas, Richardson, Texas
75080-3021, USA}
\affiliation{Center for Quantum Information, IIIS, Tsinghua University, Beijing 100084,
PR China}
\author{Chuanwei Zhang}
\thanks{Corresponding author. \\
Email: \href{mailto:chuanwei.zhang@utdallas.edu}{chuanwei.zhang@utdallas.edu}%
}
\affiliation{Department of Physics, The University of Texas at Dallas, Richardson, Texas
75080-3021, USA}

\begin{abstract}
Time-reversal invariance plays a crucial role for many exotic quantum
phases, particularly for topologically nontrivial states, in spin-orbit
coupled electronic systems. Recently realized spin-orbit coupled cold-atom
systems, however, lack the time-reversal symmetry due to the inevitable
presence of an effective transverse Zeeman field. We address this issue by
analyzing a realistic scheme to preserve time-reversal symmetry in
spin-orbit coupled ultracold atoms, with the use of Hermite-Gaussian-laser
induced Raman transitions that preserve spin-layer time-reversal symmetry.
We find that the system's quantum states form Kramers pairs, resulting in
symmetry-protected gap closing of the lowest two bands at arbitrarily large
Raman coupling. We also show that Bose gases in this setup exhibit
interaction-induced layer-stripe and uniform phases as well as intriguing
spin-layer symmetry and spin-layer correlation.
\end{abstract}

\maketitle

\section{Introduction}

\label{sec:introduction}Time-reversal invariance constitutes a fundamental
symmetry in quantum physics. A half-integer spin system always possesses
two-fold degenerate quantum states, or Kramers degeneracy~\cite{Kramers1930}%
, under the time-reversal symmetry. In solid-state materials, the presence
of time-reversal symmetry and spin-orbit coupling, interaction between
particle spin and orbital degrees of freedom, is responsible for many exotic
phenomena such as quantum spin Hall effects and topological insulators~\cite%
{Xiao2010,Qi2011,Chiu2016}, whose key physical features---a gapless edge
state---is guaranteed by Kramers degeneracy. Recently, a class of exotic
quantum phases has been found in ultracold atoms through the engineering of
various types of spin-orbit coupling via light-matter interaction~\cite%
{Higbie2002,Spielman2009,Dalibard2011,Galitski2013,Goldman2014,Zhai2015},
including spin-linear-momentum coupling~\cite%
{Stanescu2008,Wang2010,Wu2011,Ho2011,Zhang2012a,Hu2012,Ozawa2012,Li2012,Gong2011,Hu2011,Yu2011,Qu2013b,Zhang2013b,Sun2016,Martone2016,Yu2016}%
, which has been widely studied in experiments~\cite%
{Lin2011,Zhang2012b,Qu2013a,Olson2014,Hamner2014,Wang2012,Cheuk2012,Williams2013,Campbell2016,Luo2015,Huang2016,Wu2016}%
, and proposed spin-orbital-angular-momentum~\cite%
{Sun2015,Demarco2015,Qu2015,Chen2016,Jiang2016,Hou2017} as well as
spin-tensor-momentum~\cite{Luo2017} couplings. However, in these schemes,
the spin-orbit interaction is generated by a laser-induced Raman transition
between atomic hyperfine states, which manifests as a constant Zeeman field
along a fixed transverse direction and hence inevitably breaks the
time-reversal symmetry. For further pursuit of new quantum phases with
nontrivial physics due to the interplay between time-reversal symmetry and
spin-orbit couplings, it is crucial to create time-reversal invariance
coexisting with spin-orbit coupling in these systems.

\begin{figure}[b]
\centering
\includegraphics[width=1\columnwidth]{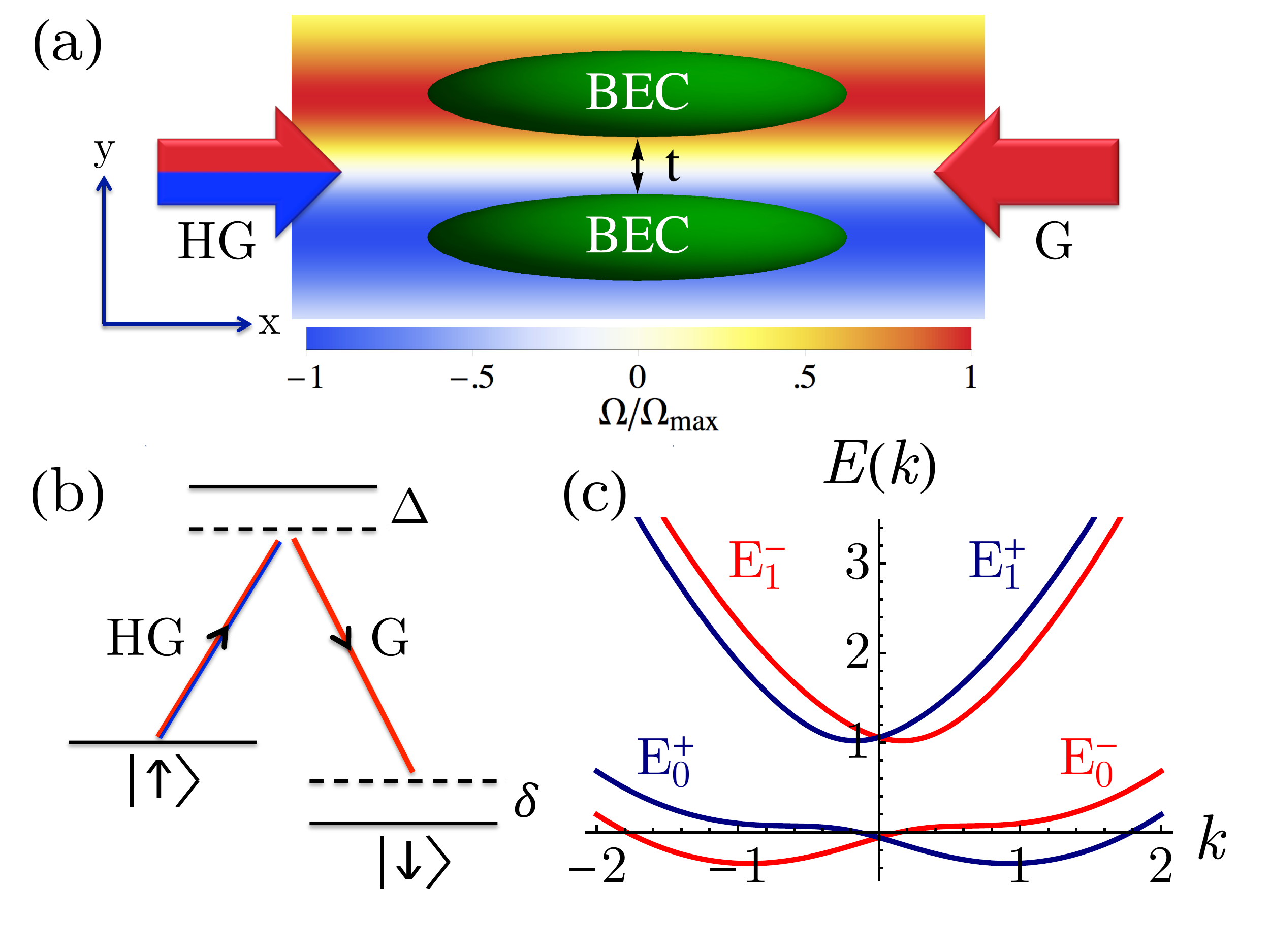} 
\caption{(a) Scheme for generating spin-orbit coupled BECs preserving
time-reversal symmetry. A pair of counter-propagating Gaussian (G) and
Hermite-Gaussian (HG) beams induces a Raman transition that has opposite
amplitudes between upper ($y>0$, red) and lower ($y<0$, blue) regions.
An additional far-detuned HG laser beam can be used to create a bilayer
structure with interlayer tunneling $t$. (b) The Raman transition process
between the atomic hyperfine levels. (c) The single-particle energy
dispersion of the system at $\Omega =1$ and $t=0.5$. The band crossing at $%
k=0$ is time-reversal symmetry protected. Each band also satisfies a
spin-layer symmetry $\langle \protect\sigma _{z}\protect\tau _{x}\rangle
=\pm 1$ (light red and dark blue, respectively). }
\label{fig:Fig_1}
\end{figure}

Ongoing study has focused on the topological structure of
spin-orbit coupled degenerate Fermi gases preserving time-reversal
symmetry~\cite{Hu2017}. In this paper, we provide detailed
analysis on such a scheme that generates time-reversal symmetry in
ultracold atoms and investigate interacting Bose-Einstein
condensates (BECs) realized with it. The approach generalizes the
conventional experimental scheme~\cite{Lin2011}, in which two
Gaussian lasers are applied to spinor gases, by replacing one
laser beam with a
first-order Hermite-Gaussian (HG) beam~\cite{Meyrath2005} [see Fig.~\ref%
{fig:Fig_1} (a)]. The HG beam induces spin-orbit coupling in each layer and
preserves the bilayer system under time-reversal operation,
\begin{equation}
\Theta =i\sigma _{y}\tau _{x}K,  \label{eq:TR_op}
\end{equation}%
where $\bm{\sigma}$ and $\bm{\tau}$ are Pauli matrices in spin and layer
space, respectively, and $K$ is the conjugate operator. Note that $\Theta $,
though involving the layer degrees of freedom, exhibits the same physical
properties as the regular time-reversal operator by being anti-unitary and $%
\Theta ^{2}=-1$. While this symmetry is crucial for nontrivial topological
states in Fermi-gas~\cite{Hu2017} and solid-state~\cite{Klinovaja2014}
systems, here, we focus on interacting Bose gases in this setup. We find
rich phase diagrams as well as interesting correlations that are not present
in regular spin-orbit coupled BECs without this time-reversal symmetry. Our
main results are summarized as below:

(i) The system's single-particle energy bands pair as time-reversal partners
and are also subject to a spin-layer symmetry $\sigma_z \tau_x$. The Kramers
degeneracy prevents the gap opening between the paired two bands at zero
momentum, resulting in double finite-momentum band minima that always exist,
i.e., the zero-momentum state can never be the single-particle ground state
even at a large Raman coupling, unless the two layers completely decouple.

(ii) The interacting phase diagram of a ground-state Bose gas exhibits
layer-stripe, plane-wave, and zero-momentum phases.
The layer-stripe phase (and its Kramers partner),
occurring at weak Raman coupling, exhibits spatially modulating
layer polarization but no total density modulation due to the
time-reversal symmetry. At large interaction and large Raman
coupling, the many-body effects drive the BEC to a zero-momentum
ground state (or its Kramers partner), even if the zero momentum
is not the single-particle band minimum.

(iii) The Bose gas exhibits a global spin-layer correlation $\langle
\sigma_x \tau_z \rangle \neq 0$, while either the spin or the layer
component vanishes, $\langle \sigma_x \rangle = \langle \tau_z \rangle = 0$.
It is crucial for experiments to measure spin and layer properties
simultaneously rather separately.

We present the details of (i) in Sec.~\ref{sec:model} and those of (ii) and
(iii) in Sec.~\ref{sec:interacting_BEC}, followed by experimental considerations in Sec.~\ref{sec:experimental consideration}
and a conclusion in Sec.~\ref{sec:conclusion}.

\section{Model and Hamiltonian}

\label{sec:model}

We start with ultracold atoms with two hyperfine spin states $({%
\begin{array}{cc}
{{\psi _{\uparrow }}} & {{\psi _{\downarrow }}}%
\end{array}%
})^{T}$, subject to a pair of counter-propagating lasers
$\Omega_L^{\pm }$ along the $x$-direction, with transverse
electromagnetic modes of a general Hermite form. The laser
amplitudes are
\[
\Omega _{m,n}^{\pm }=AH_{m}(\frac{\sqrt{2}y}{w})H_{n}(\frac{\sqrt{2}z}{w})e^{%
{-}\frac{y^{2}{+}z^{2}}{w^{2}}{\pm }ik_{R}x},
\]%
where $A$ represents the overall beam strength, $H_{n}$ is the $n$th Hermite
polynomials, $w$ is the beam waist, and $k_{R}$ is the wave vector. The two
beams impart spin-dependent linear momentum into the atoms and also induce
Raman transition $\Omega _{\mathrm{R}}(\bm{r})=\Omega ^{+\ast }\Omega
^{-}/\Delta $ between the spin states, as shown in Fig.~\ref{fig:Fig_1}(b) ($%
\Delta $ is a uniform detuning). If both beams are of the lowest mode $%
\Omega _{0,0}^{\pm }$ (two Gaussian beams), we have the regular setup for
generating the spin-linear-momentum coupling. Here, we focus on a practical
generalization to $\Omega _{1,0}^{+}$ and $\Omega _{0,0}^{-}$, i.e.,
left-propagating Gaussian and right-propagating HG beams, as shown in Fig.~%
\ref{fig:Fig_1}(a). The Raman transition amplitude $\Omega _{\mathrm{R}}(%
\bm{r})$ now has an odd spatial parity along the $y$ direction
with maximum strength at $y=\pm w/\sqrt{2}$. Along the $y$ direction, a
bilayer trapping potential can be realized using a repulsive
potential at the center of a tight harmonic trap or a single
far-detuning HG laser beam of the (1,0) mode. Such additional
far-detuning trapping lasers avoid the heating from the trapping
and ensure the independent tunability of the interlayer tunneling.
Performing a unitary transformation $\psi _{\uparrow ,\downarrow
}\rightarrow \psi _{\uparrow ,\downarrow }e^{\mp ik_{R}x}$ and
integrating out the $y$ and $z$ degrees of freedom, we obtain the
effective Hamiltonian
for upper and lower layers, respectively, as $\frac{1}{2}(p_{x}^{2}-2p_{x}%
\sigma _{z}\pm \Omega \sigma _{x})$, where $p_{x}\sigma _{z}$ is the
spin-orbit coupling and $\Omega $ is the effective Raman coupling. Here we
take $k_{R}$ and $\hbar ^{2}k_{R}^{2}/2m$ as momentum and energy units,
respectively ($m$ is the atomic mass). The only difference between the two
layers is the opposite sign of Raman coupling due to the HG beam. If the two
layers have a slight overlap, the dominant interlayer coupling is particle
tunneling between the two layers. We can treat the two layers as another
two-level degrees of freedom and write down the whole Hamiltonian in
spin-layer basis $({%
\begin{array}{cccc}
{{\psi _{1\uparrow }}} & {{\psi _{1\downarrow }}} & {{\psi _{2\uparrow }}} &
{{\psi _{2\downarrow }}}%
\end{array}%
})^{T}$, as
\begin{equation}
H=\frac{1}{2}(p_{x}^{2}-2p_{x}\sigma _{z}+\Omega \sigma _{x}\tau _{z}-t\tau
_{x}),  \label{eq:Hamiltonian}
\end{equation}%
where $t$ is the tunneling strength. Note that the
system in general can have a detuning term $\delta \sigma _{z}$
[as shown in Fig.~\ref{fig:Fig_1}(b)], which has to be tuned to
zero for the time-reversal symmetry we are interested in.

The Hamiltonian exhibits a spin-layer time-reversal symmetry,
\begin{equation}
\Theta H(p_{x})\Theta ^{-1}=H(-p_{x}),  \label{eq:TR_symmetry}
\end{equation}%
under which the lower two energy bands,
\begin{equation}
E_{0}^{\pm }(k)=\frac{1}{2}(k^{2}-\sqrt{(2k\pm t)^{2}+\Omega ^{2}}),
\label{eq:spectrum}
\end{equation}%
become time-reversal partners, i.e., $E_{0}^{+}(k)$ and
$E_{0}^{-}(-k)$ form degenerate Kramers pairs, as shown in
Fig.~\ref{fig:Fig_1}(c). The fact
$E_{0}^{+}(0)=E_{0}^{-}(0)$ leads to a symmetry protected band
crossing (or gap closing) at $k=0$. Similarly, the upper two bands
$E_{1}^{\pm }(k)=\frac{1}{2}(k^{2}+\sqrt{(2k\pm t)^{2}+\Omega
^{2}})$ are also time-reversal partners crossing at $k=0$. At
$\Omega =0$, the energy band exhibits double minima at $ \pm
k_{\mathrm{min}}= \pm 1$. As $\Omega $ increases, $k_{\rm{min}}$
shifts toward the zero momentum and approaches
$k_{\mathrm{min}}=\frac{t}{\Omega -2}+O(\Omega ^{-3})$ in the
large $\Omega $ limit ($\Omega \gg 2 $). As a result, the
single-particle ground states are doubly degenerate and always
possess finite momentum $\pm k_{\mathrm{min}}\neq 0$---the
crossing point $k=0$ can never be the ground state---with the
presence of
interlayer coupling $t\neq 0$. If the layers are completely decoupled, or $%
t=0$, the lower bands $E_{0}^{\pm }(k)$ become identical, and the ground
states undergo a transition from finite to zero momentum at $\Omega _{c}=2$,
the same critical value as in the conventional spin-orbit coupled system.

In addition to the time-reversal symmetry, the Hamiltonian also exhibits a
spin-layer symmetry,
\begin{eqnarray}  \label{eq:spin-layer_symmetry}
[H,\sigma_z\tau_x]=0.
\end{eqnarray}
We find that the paired bands $E^\mp_{0}$ (or $E^\mp_1$) are subject to $%
\langle \sigma_z\tau_x \rangle= \pm 1$ [light red and dark blue colors in Fig.~\ref%
{fig:Fig_1}(c), respectively]. By measuring this symmetry, one could
distinguish a state from its Kramers partner.

\section{Interacting Bose Gases} \label{sec:interacting_BEC}
We consider an interacting Bose gas in
this time-reversal-invariant setup and study its ground-state
properties as a function of system parameters. We use two
complementary methods, variational analysis and Gross-Pitaeviskii
equation (GPE) numerics, to find the BEC ground-state
wavefunction. Both methods show fair agreement on the results
presented in this section.

We adopt a variational wavefunction as a general superposition of a Kramers
pair as
\begin{eqnarray}  \label{eq:ansatz}
&& \Psi =\sqrt{\rho}\left[ |C_1| \left(
\begin{array}{c}
\cos\theta\cos\gamma_1 e^{i\delta_1} \\
\cos\theta\sin\gamma_1 e^{i\delta_2} \\
\sin\theta\cos\gamma_2 e^{i\delta_3} \\
\sin\theta \sin\gamma_2%
\end{array}
\right) e^{i k_1 x} \right.  \nonumber \\
&&\ \ \ \ \ \ \ \ \ \ \ \ + \left. |C_2| \left(
\begin{array}{c}
\sin\theta\sin\gamma_2 \\
-\sin\theta\cos\gamma_2 e^{-i\delta_3} \\
\cos\theta\sin\gamma_1 e^{-i\delta_2} \\
- \cos\theta \cos\gamma_1 e^{-i\delta_1}%
\end{array}
\right) e^{-i k_1 x} \right],
\end{eqnarray}
with particle number density $\rho$ and normalization condition $%
|C_1|^2+|C_2|^2=1$. The ansatz is generalized from the conventional
spin-orbit-coupled system but respects the time-reversal associated
degeneracy. By setting $\theta = \pi/4$ and $\gamma_1=\gamma_2$, the
top-layer components of $\Psi$ reproduce the previous results without
time-reversal symmetry in Ref.~\cite{Li2012}. Note that the ground state
wavefunction of this type is also confirmed by our GPE calculations, which
are not bound to any constraint.

The BEC's energy density is expressed as
\begin{eqnarray}  \label{eq:energy}
\varepsilon = \int dx \left[\Psi^\dag H \Psi + \frac{g}{2}%
|\Psi|^4+g_{\uparrow\downarrow}\sum_{j=1,2}|\psi_{j
\uparrow}|^2|\psi_{j\downarrow}|^2 \right],
\end{eqnarray}
where $g$ and $g_{\uparrow\downarrow}$ are interatomic interactions between
same and opposite spin species, respectively. Inserting Eq.~(\ref{eq:ansatz}%
) into Eq.~(\ref{eq:energy}), we obtain the energy density as a functional
of eight independent variables $k_1$, $|C_1|$, $\theta$, $\gamma_{1,2}$, and $%
\delta_{1,2,3}$ (see the Appendix). Minimizing the
functional with respect to the variables, we obtain the ground-state
wave function. We remark that $\Theta \Psi$, the time-reversal state of $\Psi$%
, is always orthogonal to $\Psi$ and gives the same energy functional $%
\varepsilon$. This means that the ground states are always doubly degenerate
and are time-reversal partners of each other. The variational ansatz also
allows us to compute the associated physical properties as
\begin{eqnarray}
\langle\sigma_z\rangle &=&(\cos^2\theta\cos 2\gamma_1+ \sin^2\theta\cos
2\gamma_2) (|C_1|^2-|C_2|^2),  \nonumber \\
\langle\sigma_x\rangle &=& [\sin2\gamma_1\cos^2\theta\cos\delta_{12}
+\sin2\gamma_2\sin^2\theta\cos\delta_3]  \nonumber \\
&&\times(|C_1|^2-|C_2|^2),  \nonumber \\
\langle\sigma_x\tau_z\rangle &=& \cos^2\theta\sin2\gamma_1\cos\delta_{12}
-\sin^2\theta\sin2\gamma_2\cos\delta_3,  \nonumber \\
\langle\sigma_z\tau_x\rangle&=& {\sin 2\theta}[\cos\gamma_1\cos\gamma_2\cos%
\delta_{13} -\sin\gamma_1\sin\gamma_2\cos\delta_2]  \nonumber \\
&&\times(|C_1|^2-|C_2|^2),
\end{eqnarray}
where $\delta_{ij} \equiv \delta_i-\delta_j$.

The interacting ground state exhibits three phases: (I) layer-stripe phase having $%
k_1 \neq 0$, $|C_1|=|C_2|=\frac{1}{\sqrt{2}}$, and $\langle\sigma_z\rangle=%
\langle\sigma_x\rangle=\langle\sigma_z \tau_x\rangle=0$
(resulting in spatially modulated layer
polarization as we will show in Sec.~\ref{sec:interacting_BEC});
(II) plane-wave
phase having $k_1 \neq 0$, $|C_1 C_2|=0$, $|\langle\sigma_z\rangle|>0$, and $%
|\langle\sigma_z \tau_x\rangle|=1$; (III) zero-momentum phase having $%
k_1=|\langle\sigma_z\rangle|=0$. Below we present the phase diagram and
discuss the physics in details.

\begin{figure}[t]
\centering
\includegraphics[width=1\columnwidth]{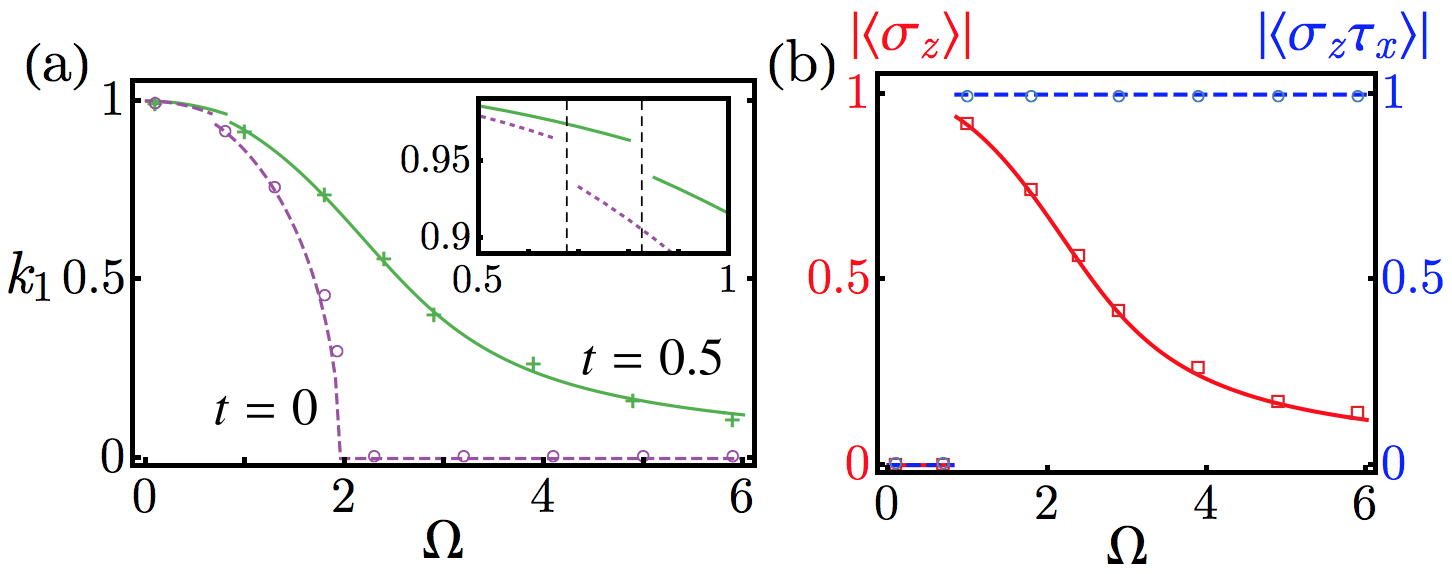} 
\caption{(a) Momentum $k_1$ vs Raman strength $\Omega$ for cases
of decoupled layers $t=0$ (purple dashed) and coupled layers $t=0.5$
(green solid). The discontinuity in both curves
(enlarged in the inset) indicates the transition
between layer-stripe and plane-wave phases. In the $t=0$ case, the
plane-wave phase can make a transition to the zero-momentum phase
($k_1$ dropping to zero), which does not occur at $t=0.5$. (b)
Spin polarization $|\langle\protect\sigma_z\rangle|
$ (red solid) and spin-layer symmetry $|\langle\protect\sigma_z\protect\tau%
_x\rangle|$ (blue dashed) vs $\Omega$ at $t=0.5$. Note that both curves are zero
and hence overlap each other in the layer-stripe phase region at small $\Omega$.
In both (a) and (b), the interaction is set to $(g,g_{\uparrow%
\downarrow})=(1,0.9)$, and curves (symbols) represent the variational
(numerical GPE) results.}
\label{fig:Fig_2}
\end{figure}

We first look at typical phase transitions for moderately interacting BEC $%
(g,g_{\uparrow \downarrow })=(1,0.9)$ as the Raman strength $\Omega $
varies. Figure \ref{fig:Fig_2}(a) shows the momentum $k_{1}$ as a
monotonically decreasing function of $\Omega $ for either coupled ($t=0.5$,
solid green curve) or decoupled ($t=0$, dashed purple curve) layers. At small $\Omega $,
the system is in the layer-stripe phase (I). As $\Omega $ increases, the $k_{1}$
curves exhibit discontinuities, at which the system undergoes a first-order
phase transition to the plane-wave phase (II). With further increase in $%
\Omega $, the $k_{1}$ curve of decoupled layers drops to zero at a critical
value, representing a second-order transition to the zero-momentum phase
(III), while that of coupled layers smoothly decreases but does not drops to
zero, i.e., no transition to phase (III). The disappearance of phase (III)
due to the interlayer coupling agrees with the single-particle physics we
discuss above.

Figure \ref{fig:Fig_2}(b) shows spin polarization $|\langle\sigma_z\rangle|$
and spin-layer symmetry $|\langle\sigma_z\tau_x\rangle|$ as a function of $%
\Omega$ for the $t=0.5$ case. We see that the layer-stripe phase (I) is spin
unpolarized $|\langle\sigma_z\rangle| = 0 $, while the plane-wave phase (II)
is spin polarized, so the big jump in spin polarization provides a good
measurable signature for the (I)--(II) transition. The spin-layer symmetry $%
|\langle\sigma_z\tau_x\rangle| = 0$ in the layer-stripe phase indicates an
interaction induced symmetry breaking that equally mixes two states of
opposite symmetry. In both Figs. \ref{fig:Fig_2}(a) and \ref{fig:Fig_2}(b), the results of
variational (curves) and GPE (symbols) calculations agree with each other.

\begin{figure}[t]
\centering
\includegraphics[width=1\columnwidth]{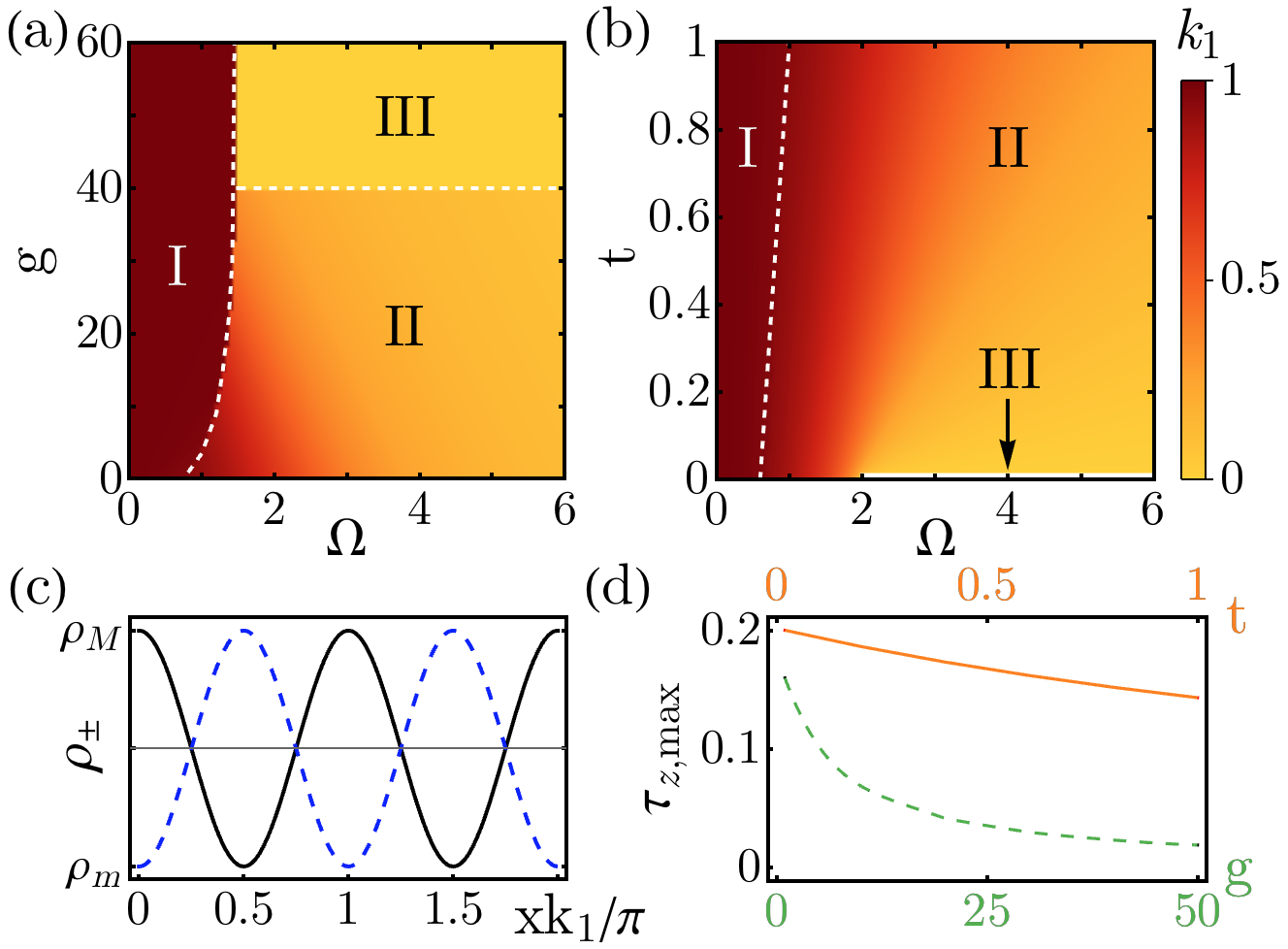} 
\caption{(a) Phase diagram in the $g$--$\Omega$ plane for $t=0.5$. In $g<40$
($>40$), the system undergoes a transition from the layer-stripe phase (I) to the
plane-wave phase (II) [zero-momentum phase (III)]. (b) Phase diagram in the $%
t$--$\Omega$ plane for $g=1$. The zero-momentum phase only occurs in the
case of decoupled layers $t=0$ at $\Omega>2$. The color in both (a) and (b)
represents momentum $k_1$, as scaled in the bar graph. (c) Top (black solid) and
bottom (blue dashed) layer density profiles $\protect\rho_{\pm}(x)$ of a layer-stripe
phase at $\Omega=t=0.5$ in (b), exhibiting out-of-phase modulations between
maximum $\protect\rho_M$ and minimum $\protect\rho_m$, or spatial
modulations in $\langle \protect\tau_z \rangle$. (d) Modulation amplitude $%
\protect\tau_{z,\mathrm{max}}$ [$\equiv (\protect\rho_M-\protect\rho_m)/2$]
versus $t$ (orange solid, top axis) and $g$ (green dashed, bottom axis). The interspin
interaction is set to be $g_{\uparrow\downarrow}=0.9g$ for all the panels.}
\label{fig:Fig_3}
\end{figure}

We turn to explore the interacting ground-state phases in a wide
parameter region. Figure \ref{fig:Fig_3}(a) shows the ground-state
phase diagram in the $g$--$\Omega$ plane for $t=0.5$ and
$g_{\uparrow\downarrow}=0.9 g$. We see that the (I)--(II) phase
transition is allowed in $0<g<40$, in which the layer-stripe phase
region increases with $g$. In $g>40$, the plane-wave phase (II)
disappears, and the system make transitions from layer-stripe (I) to
zero-momentum (III) phases. Since the zero-momentum state is never
energetically favored by the single-particle Hamiltonian with
finite $t$, the zero-momentum phase here is fully attributed to
the interaction effect. In fact, the system staying at the
zero-momentum state costs higher single-particle energy but saves
more ferromagnetic interaction energy $\propto
(g-g_{\uparrow\downarrow})\rho $. Experimentally, this region can
be achieved by increasing the atomic two-body scattering length
through the Feshbach resonance~\cite{Chin10} as well as the atomic
density.

Figure \ref{fig:Fig_3}(a) shows similarities and
essential differences compared with the phase diagram of
conventional spin-orbit coupled systems (e.g., as in
Ref.~\cite{Li2012}). While both cases exhibit the three phases and
the transition from stripe (I) to plane-wave (II) phases as
$\Omega$ increases, our system does not have the transition from
plane-wave (II) to zero-momentum (III) phases as $\Omega$ further
increases (at any nonzero $t$). Since the zero-momentum ground
state (III) is never favored on the single-particle band of our
system, its appearance in the interacting phase diagram is purely
driven by (strong) interaction. Note that in the conventional
system, the transition between plane-wave (II) to zero-momentum
(III) states already exists in the single-particle case, while the
interaction shifts the (II)--(III) phase boundary but does not
make any qualitative change.

Additionally, we present in Fig.~\ref{fig:Fig_3}%
(b) the phase diagram in the $t$-$\Omega$ plane for fixed $%
(g,g_{\uparrow\downarrow})=(1,0.9)$. We see that the interlayer tunneling
linearly increases the layer-stripe phase region with respect to $\Omega$. In the
plane-wave phase (II), the system momentum increases with $t$ at given $%
\Omega$. The zero-momentum phase (III) appears only at $t=0$, in which the
system returns to the conventional spin-orbit coupled BEC and hence exhibits
the (II)--(III) transition as shown in Fig.~\ref{fig:Fig_2}(a).

\begin{figure}[t]
\centering
\includegraphics[width=1\columnwidth]{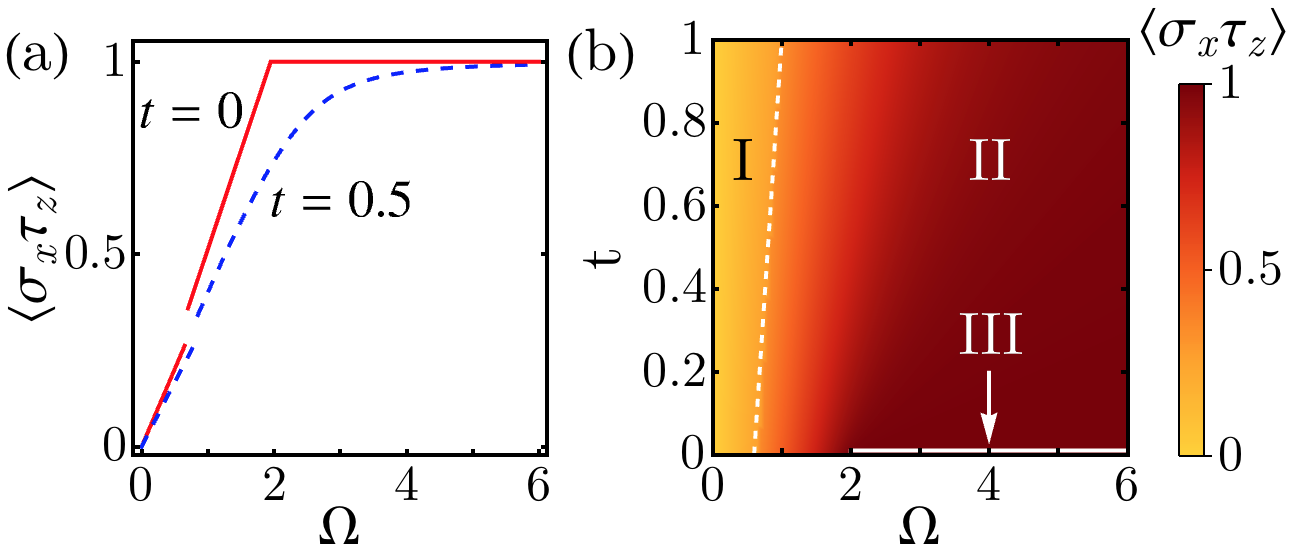} 
\caption{ (a) Spin-layer correlation $\langle\protect\sigma_x\protect\tau%
_z\rangle$ vs. Raman strength $\Omega$ in the case of decoupled layers $t=0$
(red solid) and coupled layers $t=0.5$ (blue dashed). (b) Phase diagram in the $t$--$%
\Omega$ plane with color scale representing $\langle\protect\sigma_x\protect%
\tau_z\rangle$. In both (a) and (b), the interaction is set to $%
(g,g_{\uparrow\downarrow})=(1,0.9)$. }
\label{fig:Fig_4}
\end{figure}

We further look into the detailed structure of the layer-stripe phase. Figure \ref%
{fig:Fig_3}(c) shows the top and bottom layer density profiles,
respectively, of a layer-stripe state $\Psi_s$ at $\Omega=t=0.5$ and $%
(g,g_{\uparrow\downarrow})=(1,0.9)$. The two profiles show
out-of-phase spatial modulations, or in other words, the system
exhibits no total density modulation but a
layer-polarization stripe pattern $\tau_z(x)=\Psi_s^*(x) \tau_z
\Psi_s(x)$ with wavelength $\lambda_s=\pi/k_1$. The zero total
density modulation is a direct consequence of time-reversal
symmetry, which makes the plane-wave states of $k$ and $-k$ in
Eq.~(\ref{eq:ansatz}) orthogonal to each other at any spatial
point. Note that in the conventional system of Ref.~\cite{Li2012}
or a bilayer system with same-sign Raman coupling on both layers,
the stripe phase exhibits total density modulations. Moreover,
our layer-stripe phase is doubly degenerate. The
other degenerate state is the Kramer partner $\Theta\Psi_s$, which
has a
time reversed layer-polarization modulation $-\tau_z(x)$. Figure \ref%
{fig:Fig_3}(d) shows the oscillation amplitude of $\tau_z(x)$, denoted by $%
\tau_{z,\mathrm{max}}$, vs tunneling strength $t$ and interaction
strength $g $. We see that this quantity decreases as either $t$
or $g$ increases.

Finally, we reveal an intrinsic spin-layer correlation $\langle \sigma
_{x}\tau _{z}\rangle $ of the bilayer BEC. We notice that all three phases
of the system are unpolarized in the $x$-spin direction $\langle \sigma
_{x}\rangle =0$ and density balanced between the two layers $\langle \tau
_{z}\rangle =0$. However, the product observable of both quantities $\langle
\sigma _{x}\tau _{z}\rangle $ does exhibit a non-zero expectation value. Note
that when two layers have the same $\Omega $, $\langle \sigma _{x}\tau
_{z}\rangle $ is always zero. In Fig.~\ref{fig:Fig_4}(a) we plot $\langle
\sigma _{x}\tau _{z}\rangle $ vs $\Omega $ for $t=0$ and $0.5$ at $%
(g,g_{\uparrow \downarrow })=(1,0.9)$. As $\Omega $ increases there is a
gradual increase in the spin-layer correlation, along with a discontinuity
in the curves that represents a (I)--(II) phase transition for both coupled
and decoupled layers. When our system is in the zero-momentum phase we see
that $\langle \sigma _{x}\tau _{z}\rangle =1$, which is only the case for $%
t=0$ at such low interaction strengths. Figure \ref{fig:Fig_4}(b) shows the
ground-state phase diagram in the $t$--$\Omega $ plane with color denoting $%
\langle \sigma _{x}\tau _{z}\rangle $. The layer-stripe phase has
relatively small correlations compared with the plane-wave phase.
Such a spin-layer correlation is particularly useful for
characterizing the phase diagram in experiments.

\section{Experimental consideration}
\label{sec:experimental consideration}
The experimental conditions for our time-reversal
invariant system are similar to but slightly modified from the
current setups. We consider a $^{87}$Rb bilayer BEC created in a
quasi-one-dimensional harmonic trap in the $x$ direction and a
double-well-shaped potential in the $y$ direction. The wavelength
of the HG Raman laser is 788 nm, which corresponds to the recoil
energy $E_R = h \times 3697$ Hz. If one tunes the bilayer
separation to be $0.4$ $\mu$m and the double-well barrier to be $2
E_R$, the effective tunneling is $t = 0.2 E_R$, with which
important physics of our model can be explored. Higher tunneling
is achievable by decreasing the separation or the barrier.

The layer-stripe phase can be detected by the standard
time-of-fight (TOF) image, showing the co-occupation of both
momentum minima. The out-of-phase stripe pattern may be probed
with the Bragg diffraction, which has successfully revealed the
density stripe structure of ultracold atomic gases~\cite{Li2017}.
Moreover, we suggest the use of spin-layer correlation $\langle
\sigma _{x}\tau _{z}\rangle$ to experimentally determine the phase
diagram and phase transitions. In experiment, such correlation
must be determined by performing both spin-resolved and
layer-resolved measurements. Simply measuring spin or layer
components leads to trivial results.

\section{Conclusion}

\label{sec:conclusion}  We have investigated a realistic setup for
preserving time-reversal symmetry in spin-orbit-coupled quantum
gases. Our approach has generalized the conventional setup to the
use of HG-beam induced Raman transitions, which create spin-orbit
coupling with opposite Zeeman fields for a bilayer quantum gas. As
a result, the system preserves time-reversal symmetry as well as
spin-layer symmetry. We have found that unlike the conventional
system, the time-reversal symmetry in our system leads to paired
single-particle bands due to Kramers degeneracy, which prevents
the gap opening between the lowest two bands and also the
transition from double-minimum to single-minimum band structures
as the Raman coupling increases.

We have investigated interacting Bose-Einstein condensates in this
time-reversal invariant bilayer setup. Our results show that the
ground-state phase diagram exhibits interaction-induced
layer-stripe and zero-momentum phases that can
not be characterized by the single-particle physics, as well as a
plane-wave phase analogous to the single-particle state. The
layer-stripe phase exhibits spatially modulated
layer polarization. The zero-momentum phase results from a strong
ferromagnetic interaction effect that overwhelms the
single-particle energetic favor. We have identified the parameter
region for each phase and revealed salient experimental signatures
for the phase transitions. Our work allows extensive study of the
time-reversal-invariant physics in various spin-orbit-coupled
cold-atom systems, including dynamical features of the BEC and
ground-state and excitation properties of degenerate Fermi gases
as well as high-spin atomic superfluids. Another
interesting direction is to investigate the
time-reversal-invariant quantum gases with higher dimensional
spin-orbit coupling, in which rich Kramers degeneracies may
exist~\cite{Grusdt2017,Cheng2018}.

\textbf{Acknowledgments.} This work is supported by AFOSR
(FA9550-16-1-0387), NSF (PHY-1505496), and ARO (W911NF-17-1-0128). LW
acknowledges support from National Natural Science Foundation of China
(NSFC) under grant Nos. 11475037 and Fundamental Research Funds for the
Central Universities under grant No.~DUT15LK26.

\onecolumngrid


\appendix
\section*{Appendix: Variational energy functional}
\renewcommand{\theequation}{A.\arabic{equation}}
\setcounter{equation}{0}

\label{sec:functional}  Using the variational wavefunction of
Eq.~(\ref{eq:ansatz}), we find an analytical expression for the
energy functional of Eq.~(\ref{eq:energy}). For the
single-particle energy, we have
\begin{eqnarray}
\varepsilon_0&=&\frac{k_0^2}{2}+\frac{k_1^2}{2} -{k_1k_0}(\cos^2\theta\cos
2\gamma_1+\sin^2\theta\cos 2\gamma_2) +\frac {\Omega}{2} (\cos^2\theta\sin
2\gamma_1\cos\delta_{21}- \sin^2\theta\sin 2\gamma_2\cos\delta_3 )  \nonumber
\\
&&-\frac {t} {2} \sin 2\theta (\cos\gamma_1\cos\gamma_2\cos \delta_{13} +
\sin\gamma_1\sin\gamma_2\cos \delta_2).
\end{eqnarray}%
For the interaction energy, we have the intraspin interaction,
\begin{eqnarray}
\varepsilon_g&=&\frac{ng}{2} \Big[ (1-2\beta)( \cos^4\theta\cos^4\gamma_1+
\cos^4\theta\sin^4\gamma_1+\frac{3\sin^4\theta}{4}+\frac{\sin^4\theta\cos
4\gamma_2}{4})  \nonumber \\
&&+2\beta\sin^22\theta(\cos^2\gamma_2\sin^2\gamma_1+\cos^2\gamma_1\sin^2%
\gamma_2) \Big],
\end{eqnarray}
where $\beta=|C_1|^2|C_2|^2$, and the interspin interaction,
\begin{eqnarray}
\varepsilon_f&=&\frac{ng_{\uparrow\downarrow}}{4}(1-2\beta)(
\cos^4\theta\sin^22\gamma_1+\sin^4\theta\sin^22\gamma_2)+  \nonumber \\
&&{ng_{\uparrow\downarrow}}\beta\cos^2\theta\sin^2\theta\big[%
2\cos^2\gamma_1\cos^2\gamma_2+2\sin^2\gamma_1\sin^2\gamma_2-\sin
2\gamma_1\sin 2\gamma_2\cos(\delta_1-\delta_2-\delta_3)\big].
\end{eqnarray}%
\vspace{-5px} Combining these we get the full energy functional $%
\varepsilon=\varepsilon_0+\varepsilon_g+\varepsilon_f$.

\twocolumngrid


\begin{thebibliography}{99}


\bibitem{Kramers1930} H. A. Kramers, Th\'eorie g\'en\'erale de la rotation paramagn\'etique
dans les cristaux, Proc. Amsterdam Acad. \textbf{33}, 959 (1930).



\bibitem{Xiao2010} D. Xiao, M.-C. Chang, and Q. Niu, {Berry phase effects on
electronic properties}, \href{https://doi.org/10.1103/RevModPhys.82.1959}{%
Rev. Mod. Phys. \textbf{82}, 1959 (2010)}.

\bibitem{Qi2011} X.-L. Qi and S.-C. Zhang, {Topological insulators and
superconductors}, \href{https://doi.org/10.1103/RevModPhys.83.1057}{Rev.
Mod. Phys. \textbf{83}, 1057 (2011)}.

\bibitem{Chiu2016} C.-K. Chiu, J. C. Y. Teo, A. P. Schnyder, and S. Ryu, {%
Classification of topological quantum matter with symmetries}, \href{https://doi.org/10.1103/RevModPhys.88.035005}%
{Rev. Mod. Phys. \textbf{88}, 035005 (2016)}.


\bibitem{Higbie2002} J. Higbie and D. M. Stamper-Kurn, {Periodically Dressed
Bose-Einstein Condensate: A Superfluid with an Anisotropic and Variable
Critical Velocity}, \href{http://dx.doi.org/10.1103/PhysRevLett.88.090401}{%
Phys. Rev. Lett. \textbf{88}, 090401 (2002)}.

\bibitem{Spielman2009} I. B. Spielman, {Raman processes and effective gauge
potentials}, \href{http://dx.doi.org/10.1103/PhysRevA.79.063613}{Phys. Rev.
A \textbf{79}, 063613 (2009)}.

\bibitem{Dalibard2011} J. Dalibard, F. Gerbier, G. Juzeli\={u}nas, and P.
\"{O}hberg, {Colloquium: Artificial gauge potentials for neutral atoms},
\href{https://doi.org/10.1103/RevModPhys.83.1523}{Rev. Mod. Phys. \textbf{83}%
, 1523 (2011)}.

\bibitem{Galitski2013} V. Galitski and I. B. Spielman, {Spin-orbit coupling
in quantum gases}, \href{https://doi.org/10.1038/nature11841}{Nature \textbf{%
494}, 49 (2013)}.

\bibitem{Goldman2014} N. Goldman, G. Juzeli\={u}nas, P. \"{O}hberg, and I.
B. Spielman, {Light-induced gauge fields for ultracold atoms}, \href{https://doi.org/10.1088/0034-4885/77/12/126401}%
{Rep. Prog. Phys. \textbf{77}, 126401 (2014)}.

\bibitem{Zhai2015} H. Zhai, {Degenerate quantum gases with spin-orbit
coupling: a review}, \href{https://doi.org/10.1088/0034-4885/78/2/026001}{%
Rep. Prog. Phys. \textbf{78}, 026001 (2015)}.


\bibitem{Stanescu2008} T. D. Stanescu, B. Anderson, and V. Galitski, {%
Spin-orbit coupled Bose-Einstein condensates}, \href{https://doi.org/10.1103/PhysRevA.78.023616}%
{Phys. Rev. A \textbf{78}, 023616 (2008)}.

\bibitem{Wang2010} C. Wang, C. Gao, C.-M. Jian, and H. Zhai, {Spin-Orbit
Coupled Spinor Bose-Einstein Condensates}, \href{https://doi.org/10.1103/PhysRevLett.105.160403}%
{Phys. Rev. Lett. \textbf{105}, 160403 (2010)}.

\bibitem{Wu2011} C. Wu, I. Mondragon-Shem, and X.-F. Zhou, {Unconventional
Bose-Einstein Condensations from Spin-Orbit Coupling}, \href{https://doi.org/10.1088/0256-307X/28/9/097102}%
{Chin. Phys. Lett. \textbf{28}, 097102 (2011)}.

\bibitem{Ho2011} T.-L. Ho and S. Zhang, {Bose-Einstein Condensates with
Spin-Orbit Interaction}, \href{https://doi.org/10.1103/PhysRevLett.107.150403}%
{Phys. Rev. Lett. \textbf{107}, 150403 (2011)}.

\bibitem{Zhang2012a} Y. Zhang, L. Mao, and C. Zhang, {Mean-Field Dynamics of
Spin-Orbit Coupled Bose-Einstein Condensates}, \href{https://doi.org/10.1103/PhysRevLett.108.035302}%
{Phys. Rev. Lett. \textbf{108}, 035302 (2012)}.

\bibitem{Hu2012} H. Hu, B. Ramachandhran, H. Pu, and X.-J. Liu, {Spin-Orbit
Coupled Weakly Interacting Bose-Einstein Condensates in Harmonic Traps},
\href{https://doi.org/10.1103/PhysRevLett.108.010402}{Phys. Rev. Lett.
\textbf{108}, 010402 (2012)}.

\bibitem{Ozawa2012} T. Ozawa and G. Baym, {Stability of Ultracold Atomic
Bose Condensates with Rashba Spin-Orbit Coupling against Quantum and Thermal
Fluctuations}, \href{https://doi.org/10.1103/PhysRevLett.109.025301}{Phys.
Rev. Lett. \textbf{109}, 025301 (2012)}.

\bibitem{Li2012} Y. Li, L. P. Pitaevskii, and S. Stringari, {Quantum
Tricriticality and Phase Transitions in Spin-Orbit Coupled Bose-Einstein
Condensates}, \href{https://doi.org/10.1103/PhysRevLett.108.225301}{Phys.
Rev. Lett. \textbf{108}, 225301 (2012)}.

\bibitem{Gong2011} M. Gong, S. Tewari, and C. Zhang, {BCS-BEC Crossover and
Topological Phase Transition in 3D Spin-Orbit Coupled Degenerate Fermi Gases}%
, \href{https://doi.org/10.1103/PhysRevLett.107.195303}{Phys. Rev. Lett.
\textbf{107}, 195303 (2011)}.

\bibitem{Hu2011} H. Hu, L. Jiang, X.-J. Liu, and H. Pu, {Probing Anisotropic
Superfluidity in Atomic Fermi Gases with Rashba Spin-Orbit Coupling}, \href{https://doi.org/10.1103/PhysRevLett.107.195304}%
{Phys. Rev. Lett. \textbf{107}, 195304 (2011)}.

\bibitem{Yu2011} Z.-Q. Yu and H. Zhai, {Spin-Orbit Coupled Fermi Gases
across a Feshbach Resonance}, \href{https://doi.org/10.1103/PhysRevLett.107.195305}%
{Phys. Rev. Lett. \textbf{107}, 195305 (2011)}.

\bibitem{Qu2013b} C. Qu, Z. Zheng, M. Gong, Y. Xu, Li Mao, X. Zou, G. Guo,
and C Zhang, {Topological superfluids with finite-momentum pairing and
Majorana fermions}, \href{https://doi.org/10.1038/ncomms3710}{Nat. Commun.
\textbf{4}, 2710 (2013)}.

\bibitem{Zhang2013b} W. Zhang and W. Yi, {Topological
Fulde-Ferrell-Larkin-Ovchinnikov states in spin--orbit-coupled Fermi gases},
\href{https://doi.org/10.1038/ncomms3711}{Nat. Commun. \textbf{4}, 2711
(2013)}.

\bibitem{Sun2016} K. Sun, C. Qu, Y. Xu, Y. Zhang, and C. Zhang, {Interacting
spin-orbit-coupled spin-1 Bose-Einstein condensates}, \href{https://doi.org/10.1103/PhysRevA.93.023615}%
{Phys. Rev. A \textbf{93}, 023615 (2016)}.

\bibitem{Martone2016} G. Martone, F. Pepe, P. Facchi, S. Pascazio, and S.
Stringari, {Tricriticalities and Quantum Phases in Spin-Orbit-Coupled Spin-1
Bose Gases}, \href{https://doi.org/10.1103/PhysRevLett.117.125301}{Phys.
Rev. Lett. \textbf{117}, 125301 (2016)}.

\bibitem{Yu2016} Z.-Q Yu, {Phase transitions and elementary excitations in
spin-1 Bose gases with Raman-induced spin-orbit coupling}, \href{https://doi.org/10.1103/PhysRevA.93.033648}%
{Phys. Rev. A \textbf{93}, 033648 (2016)}.


\bibitem{Lin2011} Y.-J. Lin, K. Jim\'{e}nez-Garc\'{\i}a, and I. B. Spielman,
{Spin-orbit-coupled Bose-Einstein condensates}, \href{https://doi.org/10.1038/nature09887}%
{Nature (London) \textbf{471}, 83 (2011)}.

\bibitem{Zhang2012b} J.-Y. Zhang, S.-C. Ji, Z. Chen, L. Zhang, Z.-D. Du, B.
Yan, G.-S. Pan, B. Zhao, Y.-J. Deng, H. Zhai, S. Chen, and J.-W. Pan, {%
Collective Dipole Oscillations of a Spin-Orbit Coupled Bose-Einstein
Condensate}, \href{https://doi.org/10.1103/PhysRevLett.109.115301}{Phys.
Rev. Lett. \textbf{109}, 115301 (2012)}.

\bibitem{Qu2013a} C. Qu, C. Hamner, M. Gong, C. Zhang, and P. Engels, {%
Observation of Zitterbewegung in a spin-orbit-coupled Bose-Einstein
condensate}, \href{https://doi.org/10.1103/PhysRevA.88.021604}{Phys. Rev. A
\textbf{88}, 021604(R) (2013)}.

\bibitem{Olson2014} A. J. Olson, S.-J. Wang, R. J. Niffenegger, C.-H. Li, C.
H. Greene, and Y. P. Chen, {Tunable Landau-Zener transitions in a
spin-orbit-coupled Bose-Einstein condensate}, \href{https://doi.org/10.1103/PhysRevA.90.013616}%
{Phys. Rev. A \textbf{90}, 013616 (2014)}.

\bibitem{Hamner2014} C. Hamner, C. Qu, Y. Zhang, J. Chang, M. Gong, C.
Zhang, and P. Engels, {Dicke-type phase transition in a spin-orbit-coupled
Bose-Einstein condensate}, \href{https://doi.org/10.1038/ncomms5023}{Nat.
Commun. \textbf{5}, 4023 (2014)}.

\bibitem{Wang2012} P. Wang, Z.-Q. Yu, Z. Fu, J. Miao, L. Huang, S. Chai, H.
Zhai, and J. Zhang, {Spin-Orbit Coupled Degenerate Fermi Gases}, \href{https://doi.org/10.1103/PhysRevLett.109.095301}%
{Phys. Rev. Lett. \textbf{109}, 095301 (2012)}.

\bibitem{Cheuk2012} L. W. Cheuk, A. T. Sommer, Z. Hadzibabic, T. Yefsah, W.
S. Bakr, and M. W. Zwierlein, {Spin-Injection Spectroscopy of a Spin-Orbit
Coupled Fermi Gas}, \href{https://doi.org/10.1103/PhysRevLett.109.095302}{%
Phys. Rev. Lett. \textbf{109}, 095302 (2012)}.

\bibitem{Williams2013} R. A. Williams, M. C. Beeler, L. J. LeBlanc, K. Jim%
\'{e}nez-Garc\'{\i}a, and I. B. Spielman, {Raman-Induced Interactions in a
Single-Component Fermi Gas Near an $s$-Wave Feshbach Resonance}, \href{https://doi.org/10.1103/PhysRevLett.111.095301}%
{Phys. Rev. Lett. \textbf{111}, 095301 (2013)}.

\bibitem{Campbell2016} D. L. Campbell, R. M. Price, A. Putra, A.
Vald\'es-Curiel, D. Trypogeorgos, and I. B. Spielman, {Magnetic phases of
spin-1 spin-orbit-coupled Bose gases}, \href{https://doi.org/10.1038/ncomms10897}%
{Nat. Commun. \textbf{7}, 10897 (2016)}.

\bibitem{Luo2015} X. Luo, L. Wu, J. Chen, Q. Guan, K. Gao, Z.-F. Xu, L. You,
R. Wang, {Tunable spin-orbit coupling synthesized with a modulating gradient
magnetic field}, \href{https://doi.org/10.1038/srep18983}{Sci. Rep. \textbf{6%
}, 18983 (2016)}.

\bibitem{Huang2016} L. Huang, Z. Meng, P. Wang, P. Peng, S.-L. Zhang, L.
Chen, D. Li, Q. Zhou, and J. Zhang, {Experimental realization of
two-dimensional synthetic spin-orbit coupling in ultracold Fermi gases}, \href{https://doi.org/10.1038/nphys3672}%
{Nat. Phys. \textbf{12}, 540 (2016)}.

\bibitem{Wu2016} Z. Wu, L. Zhang, W. Sun, X.-T. Xu, B.-Z. Wang, S.-C. Ji, Y.
Deng, S. Chen, X.-J. Liu, and J.-W. Pan, {Realization of two-dimensional
spin-orbit coupling for Bose-Einstein condensates}, \href{https://doi.org/10.1126/science.aaf6689}%
{Science \textbf{354}, 83 (2016)}.


\bibitem{Sun2015} K. Sun, C. Qu, and C. Zhang, {%
Spin--orbital-angular-momentum coupling in Bose-Einstein condensates}, \href{https://doi.org/10.1103/PhysRevA.91.063627}%
{Phys. Rev. A \textbf{91}, 063627 (2015)}.

\bibitem{Demarco2015} M. Demarco, and H. Pu, {Angular spin-orbit coupling in
cold atoms}, \href{https://doi.org/10.1103/PhysRevA.91.033630}{Phys. Rev. A
\textbf{91}, 033630 (2015)}.

\bibitem{Qu2015} C. Qu, K. Sun, and C. Zhang, {Quantum phases of
Bose-Einstein condensates with synthetic spin--orbital-angular-momentum
coupling}, \href{https://doi.org/10.1103/PhysRevA.91.053630}{Phys. Rev. A
\textbf{91}, 053630 (2015)}.

\bibitem{Chen2016} L. Chen, H. Pu, and Y. Zhang, {Spin-orbit angular
momentum coupling in a spin-1 Bose-Einstein condensate}, \href{https://doi.org/10.1103/PhysRevA.93.013629}%
{Phys. Rev. A \textbf{93}, 013629 (2016)}.

\bibitem{Jiang2016} L. Jiang, Y. Xu, and C. Zhang, {Phase-tunable Josephson
junction and spontaneous mass current in a spin-orbit-coupled Fermi
superfluid}, \href{https://doi.org/10.1103/PhysRevA.94.043625}{Phys. Rev. A
\textbf{94}, 043625 (2016)}.

\bibitem{Hou2017} J. Hou, X.-W. Luo, K. Sun, and C. Zhang, {Adiabatically
tuning quantized supercurrents in an annular Bose-Einstein condensate},
\href{https://doi.org/10.1103/PhysRevA.96.011603}{Phys. Rev. A \textbf{96},
011603(R) (2017)}.


\bibitem{Luo2017} X.-W. Luo, K. Sun, and C. Zhang, {%
Spin-tensor--momentum-coupled Bose-Einstein condensates}, \href{https://doi.org/10.1103/PhysRevLett.119.193001}%
{Phys. Rev. Lett. \textbf{119}, 193001 (2017)}.


\bibitem{Hu2017} H. Hu, F. Zhang, and C. Zhang, {Majorana Doublets, Flat
Bands, and Dirac Nodes in s-Wave Superfluids}, \href{https://arxiv.org/abs/1710.06388}%
{arXiv:1710.06388}.


\bibitem{Meyrath2005} T. P. Meyrath, F. Schreck, J. L. Hanssen, C. -S. Chuu,
and M. G. Raizen, A high frequency optical trap for atoms using
Hermite-Gaussian beams, \href{https://doi.org/10.1364/OPEX.13.002843}{Opt.
Express \textbf{13}, 2843 (2005)}.


\bibitem{Klinovaja2014} C. Reeg, C. Schrade, J. Klinovaja, and D. Loss,
DIII topological superconductivity with emergent time-reversal symmetry, \href{https://doi.org/10.1103/PhysRevB.96.161407}%
{Phys. Rev. B \textbf{96}, 161407(R) (2017)}.


\bibitem{Chin10} C. Chin, R. Grimm, P. Julienne, and E. Tiesinga, {Feshbach
resonances in ultracold gases}, \href{https://dx.doi.org/10.1103/RevModPhys.82.1225}%
{Rev. Mod. Phys. \textbf{82}, 1225 (2010)}.

\bibitem{Li2017} J.-R. Li, J. Lee, W. Huang, S. Burchesky, B. Shteynas, F.
\c{C}. Top, A. O. Jamison, and W. Ketterle, {A stripe phase with
supersolid
properties in spin-orbit-coupled Bose-Einstein condensates}, \href{https://doi.org/10.1038/nature21431}%
{Nature (London) \textbf{543}, 91 (2017)}.

\bibitem{Grusdt2017}Tunable spin-orbit coupling for ultracold atoms in two-dimensional optical
lattices, F. Grusdt, T. Li, I. Bloch, and E. Demler,
\href{https://doi.org/10.1103/PhysRevA.95.063617}{Phys. Rev. A
\textbf{95}, 063617 (2017)}.

\bibitem{Cheng2018}Symmetry-enriched Bose-Einstein condensates in a spin-orbit-coupled bilayer system, J.-M. Cheng, X.-Fa Zhou, Z.-Wei Zhou, G.-C. Guo, and M. Gong,
\href{https://doi.org/10.1103/PhysRevA.97.013625}{Phys. Rev. A
\textbf{97}, 013625 (2018)}.



\end{thebibliography}
\end{document}